\documentclass[aps,prl,showpacs,amsmath,amssymb,twocolumn,superscriptaddress]{revtex4-1}
\pdfoutput=1

\usepackage{graphicx}
\usepackage{color}
\usepackage{amsmath}

\usepackage{soul}

\begin{document}

\title{Flat band superconductivity in strained Dirac materials}

\author{V. J. Kauppila}
\email[]{ville.kauppila@aalto.fi}
\affiliation{Low Temperature Laboratory, Department of Applied Physics, Aalto University, P.O. Box 15100, FI-00076 Aalto, Finland}
\author{F. Aikebaier}
\affiliation{University of Jyv\"askyl\"a, Department of Physics and Nanoscience Center, P.O. Box 35, FI-40014 University of Jyv\"askyl\"a, Finland}
\author{T. T. Heikkil\"a}
\affiliation{University of Jyv\"askyl\"a, Department of Physics and Nanoscience Center, P.O. Box 35, FI-40014 University of Jyv\"askyl\"a, Finland}

\newcommand{\tmpnote}[1]%
   {\begingroup{\it (FIXME: #1)}\endgroup}
   \newcommand{\comment}[1]%
       {\marginpar{\tiny C: #1}}

\date{\today}

\begin{abstract}
We consider superconducting properties of a two-dimensional Dirac material such as graphene under strain that produces a flat band spectrum in the normal state. We show that in the superconducting state, such a model results in a highly increased critical temperature compared to the case without the strain, inhomogenous order parameter with two-peak shaped local density of states and yet a large and almost uniform and isotropic supercurrent. This model could be realized in strained graphene or ultracold atom systems and could be responsible for unusually strong superconductivity observed in some graphite interfaces and certain IV-VI semiconductor heterostructures.
\end{abstract}


\maketitle


In conventional superconductors, the superconducting critical temperature $T_c$ depends exponentially on the electronic density of states $\nu$ at the Fermi level, $T_c \sim e^{-1/(g \nu)}$, where $g$ describes the strength of attractive interaction. Thus, when engineering materials for higher critical temperatures, it is natural to aim to increase the density of states. In two-dimensional systems such as graphene, a traditional approach for this is to utilize doping \cite{profeta2012phonon}, which recently lead to a breakthrough as strongly doped graphene was found to be superconducting with $T_c$ of a few Kelvin \cite{chapman2015superconductivity}. An extreme case of increased density of states is the flat band state, where the electrons within some momentum regime are completely dispersionless, leading to diverging density of states at the corresponding energy. In various different models, this has been shown to result in a parametrically enhanced critical temperature that is linear in the electron-phonon coupling constant, $T_c \sim g$ \cite{kopnin2011high,khodel1990superfluidity}. It has also been shown that this type of an approximate flat band state is realized in graphene and other Dirac materials under periodic strain \cite{tang2014strain,vozmediano2010gauge}.

Besides straining Dirac electrons, there have been several propositions for realizing systems with a flat band and possibly promoting superconductivity \cite{heikkilaup2015}. Such models include surface states of topological semimetals with an approximate chiral symmetry \cite{heikkila2011flat}. If the energy scale characterizing the deviation from the exact symmetry is weaker than that characterizing superconductivity, the mean field theory predicts flat band superconductivity \cite{kopnin2013high}. An example system belonging to this class is rhombohedral graphite. However, this type of superconductors are prone to fluctuations \cite{kauppila2016}. 

The most often encountered models leading to flat bands result from large magnetic fields and the associated Landau levels \cite{landau1930}. However,  magnetic fields also break the time reversal symmetry and typically suppress (singlet) superconducting order, so they cannot directly be utilized. A recent approach was hence to study superconductivity in a time-reversal invariant attractive Harper--Hubbard model defined on a two-dimensional square lattice \cite{peotta2015}, with the most direct realization in ultracold gases.

Here we present a BCS-like model for superconductivity of Dirac electrons under the type of strain that produces a flat band normal state. This model has quite possibly been already realized in interfaces between IV-VI semiconductor heterostructures where the strain is naturally created between a topological insulator and a trivial insulator due to lattice mismatch \cite{tang2014strain,fogel2002interfacial,fogel2001novel}. Another possible realization for this model is in graphene with a strain field, created either artificially, or at an interface inside graphite \cite{esquinazi2014}. The latter suggestion also builds on the recent experimental evidence that graphene can become superconducting under heavy doping\cite{profeta2012phonon,chapman2015superconductivity}. Our proposal could potentially be used to increase the superconducting critical temperature much higher in the absence of external doping. Besides, our model can be studied in ultracold gases in optical lattices where transforming Dirac points with adjustable geometry has already been demonstrated \cite{tarruell2012creating} and where the interaction between the electrons can be tuned via Feshbach resonances \cite{bloch2008,chin2010}.

\begin{figure}[!ht]
\centering
\includegraphics[width=0.98\columnwidth]{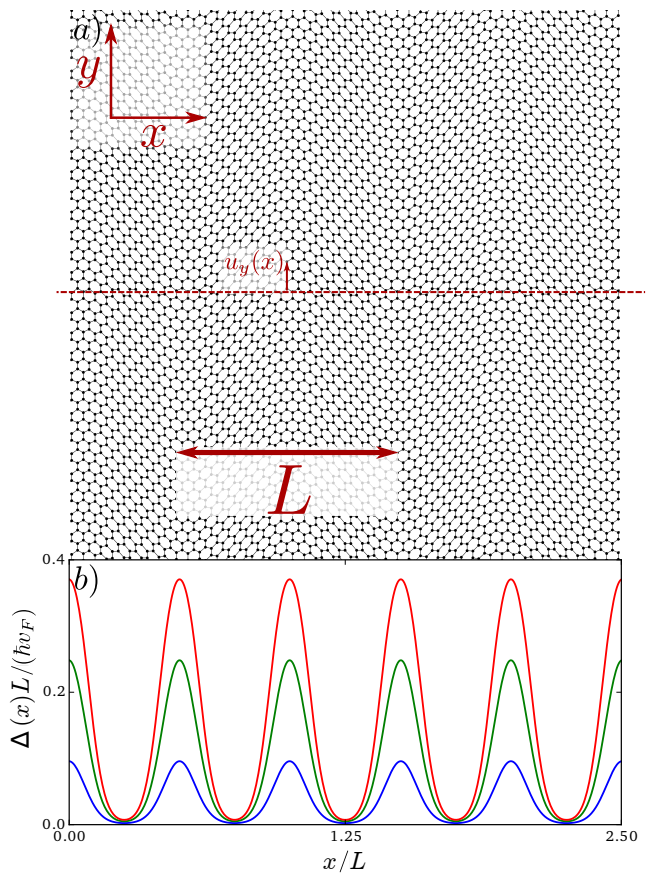}
\caption{\small a) Schematic, highly exaggerated picture of a honeycomb lattice in strain field of the form $u_y(x) = \frac{a \beta}{4 \pi} \sin (2\pi x / L)$, $L$ being the strain period. b) The profile of $\Delta(x)$ for $\beta=20$ (blue), $\beta=30$ (green) and $\beta=40$ (red) and $g / (\hbar v_F L) = 0.01$.}
\label{fig:straing}
\end{figure}

\begin{figure}[!ht]
\centering
\includegraphics[width=0.98\columnwidth]{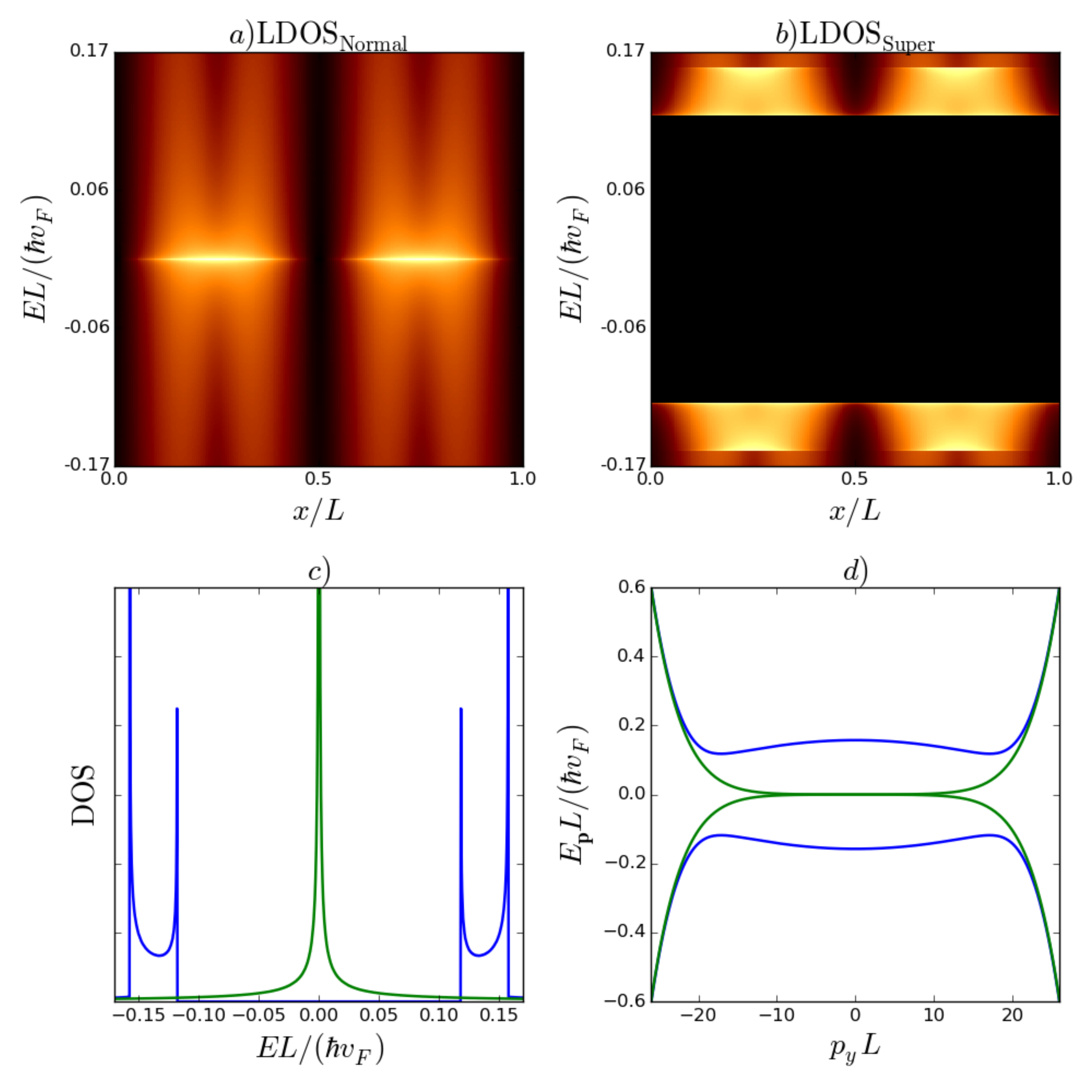}
\caption{Spectral properties of Dirac electrons in a periodic strain in normal and superconducting states with $\beta=40$ and $g / (\hbar v_F L) =0.005$. a) Local density of states in the normal state. b) Local density of states in the superconducting state. c) Total density of states for the normal (green) and superconducting (blue) states. d) The energy spectrum in both states.}
\label{fig:dos}
\end{figure}

To achieve the flat band state, the strain field experienced by the Dirac electrons should be such that the resulting effective vector potential is of the form $\mathbf{A} \propto (0, A_y(x), 0)$, where $A_y(x)$, the vector potential in the $y$-direction in the 2D lattice, changes sign periodically in $x$, the direction perpendicular to the vector potential. A strict periodical variation is not entirely necessary for this effect, but it allows for a more direct theoretical description of the effect. Here we follow \cite{tang2014strain} and use $A_y(x) = \frac{\beta}{L} \cos(2 \pi x / L)$, where $\beta$ is a dimensionless parameter describing the strength of the strain and $L$ is the strain period. In graphene, this vector potential could be produced for example by an in-plane strain field of the form $u_y(x) = \frac{a \beta}{4 \pi} \sin(2 \pi x / L)$ (assuming graphene Gr\"uneisen parameter $=2$ \cite{vozmediano2010gauge}) or out-of-plane strain field of the form $h(x,y) = y+\frac{a \beta}{4\pi} \sin( 2\pi x / L)$, where $a=1.42$\AA~is the graphene lattice constant. As a result, the low-energy Hamiltonian describing the Dirac electrons is given by
\begin{equation}
\hat{H}_0 = \hbar v_F \hat{p}_x \sigma_x + \hbar v_F \left(\hat{p}_y + A_y(x)\right) \sigma_y ,
\label{H0}
\end{equation}
where $v_F$ is the Fermi velocity of the Dirac material. In condensed-matter systems, Dirac points appear in pairs (valleys in graphene physics). Equation \eqref{H0} describes the physics at one valley, say $\mathbf{K}$, whereas the sign of $A_y$ is reversed for the partner valley ($\mathbf{K}'$). The dispersion relation of this model has an approximate flat band for $p_x \in [-\pi/L, \pi/L ]$ and $p_y \in [-\beta / (2 L), \beta / (2 L)]$. Inside the flat band, it has a weak dispersion of the form $E_{\mathbf{p}} = \hbar c \vert \mathbf{p} \vert$, where $c = v_F/I_0(\beta/ \pi)$ and $I_n(x)$ is the modified Bessel function of the first kind \cite{supplement}. For $\beta \gg \pi$, the speed becomes exponentially small, and the bands become asymptotically flat. This dispersion, along with the width of the flat band, determines the energy scale above which the model can be considered to have a flat band. The eigenstates of this Hamiltonian are localized at the points where the potential changes sign so that one sublattice is occupied at one sign change and the other sublattice at the opposite sign change. This can be seen in the density of states of the normal state shown in Fig. \ref{fig:dos}a.

The Hamiltonian \eqref{H0} is closely related to the Su-Schrieffer-Heeger (SSH) model for polyacetylene chains \cite{heeger1988solitons}. When the $y$-directional momentum $p_y$ in \eqref{H0} is zero, the model is exactly the SSH model with a domain wall and the associated topological zero energy state at each point where the potential changes sign. For finite $p_y$ the domain walls move closer to each other and the zero energy states start to overlap until they are effectively destroyed at $p_y \approx \beta / (2 L)$.

When we add an attractive interaction of coupling strength $g$ to the model \eqref{H0} (we assume here s-wave type coupling for simplicity \cite{swavenote}), the material has the possibility to enter a superconducting state described by the Bogoliubov-de Gennes equation
\begin{equation}
\begin{pmatrix}
H_0 & \Delta(x) \\
\Delta^*(x) & -H_0
\end{pmatrix}
\begin{pmatrix}
u_n(\mathbf{r}) \\
v_n(\mathbf{r})
\end{pmatrix}
=
E_n
\begin{pmatrix}
u_n(\mathbf{r}) \\
v_n(\mathbf{r})
\end{pmatrix} .
\label{hbdg}
\end{equation}
The two degrees of freedom in the matrix are the Dirac particles, described by $H_0$ given in \eqref{H0} with spinor wave functions $u_n(\mathbf{r})$, and Dirac holes, described by the Hamiltonian for holes, $-H_0$ with spinor wave functions $v_n(\mathbf{r})$. The form of the hole Hamiltonian follows from the fact that the hole partner for a particle in valley $\mathbf{K}$ is a conjugated particle at valley $-\mathbf{K}$ so that $H_0^{\mathrm{holes}} ( \mathbf{K}) = H_0^*(-\mathbf{K}) = -H_0(\mathbf{K})$. Below we suppress the subvalley indices and consider only $\mathbf{K}$ subvalley except when otherwise mentioned. The coupling between the particles and holes is described by a superconducting order parameter, $\Delta(x)$, which, because of the periodic potential in the Hamiltonian, has a periodic dependence on the $x$ coordinate. It can be found by solving the self-consistency equation
\begin{equation}
\begin{split}
\Delta(x) = & \frac{gL^2}{4 \pi^2}\int_0^{2\pi/L} d p_x \int_0^{p_c} d p_y \sum_i\sum_n \\
& \times v_{i,n,\mathbf p}^*(x)u_{i,n\mathbf p}(x) \tanh\left[\frac{E_n({\mathbf p})}{2 k_B T}\right],
\end{split}
\label{eq:selfcons}
\end{equation}
where we have summed over the two valleys which leads to both sublattices, labeled by $i$, contributing to the same $\Delta$. The sum over the band index $n$ can be restricted to those bands with energy below some cutoff energy, say the Debye energy due to the electron-phonon interaction.

With qualitative analysis of the energy scales of the model, we find three different regimes: (i) When $g \ll \hbar c L$, the small linear slope of the spectrum is visible and superconductivity is of the type found in pure Dirac materials \cite{kopnin2008bcs,lozovik2010theory} (for example, graphene). In particular, there is a quantum critical point at $g = \pi^2 \hbar c L / \beta$ below which superconductivity does not take place. (ii) When $\hbar c L \ll g \ll \hbar v_F L / \beta$, the system is in the pure flat band superconductivity regime, only the lowest energy band contributing to the superconductivity. The critical temperature is enhanced and $\Delta$ is localized. This is the interesting regime where we focus below. (iii) When $g \geq \hbar v_F L / \beta$, also higher (non-flat) bands contribute to superconductivity. A model resembling the one here in this third limit was considered in \cite{hosseini2015inhomogeneous} in strained graphene where an inhomogenous superconducting state was also found.

Because the density of states is peaked at the locations of the vector potential sign change, we expect that the superconducting order parameter is also localized close to these points. To demonstrate this, we calculate $\Delta(x)$ in the model \eqref{hbdg} from the self-consistency calculation \cite{supplement}. The resulting profile of $\Delta(x)$ is shown in Fig.~\ref{fig:straing}b.

\begin{figure}[!ht]
\centering
\includegraphics[width=0.98\columnwidth]{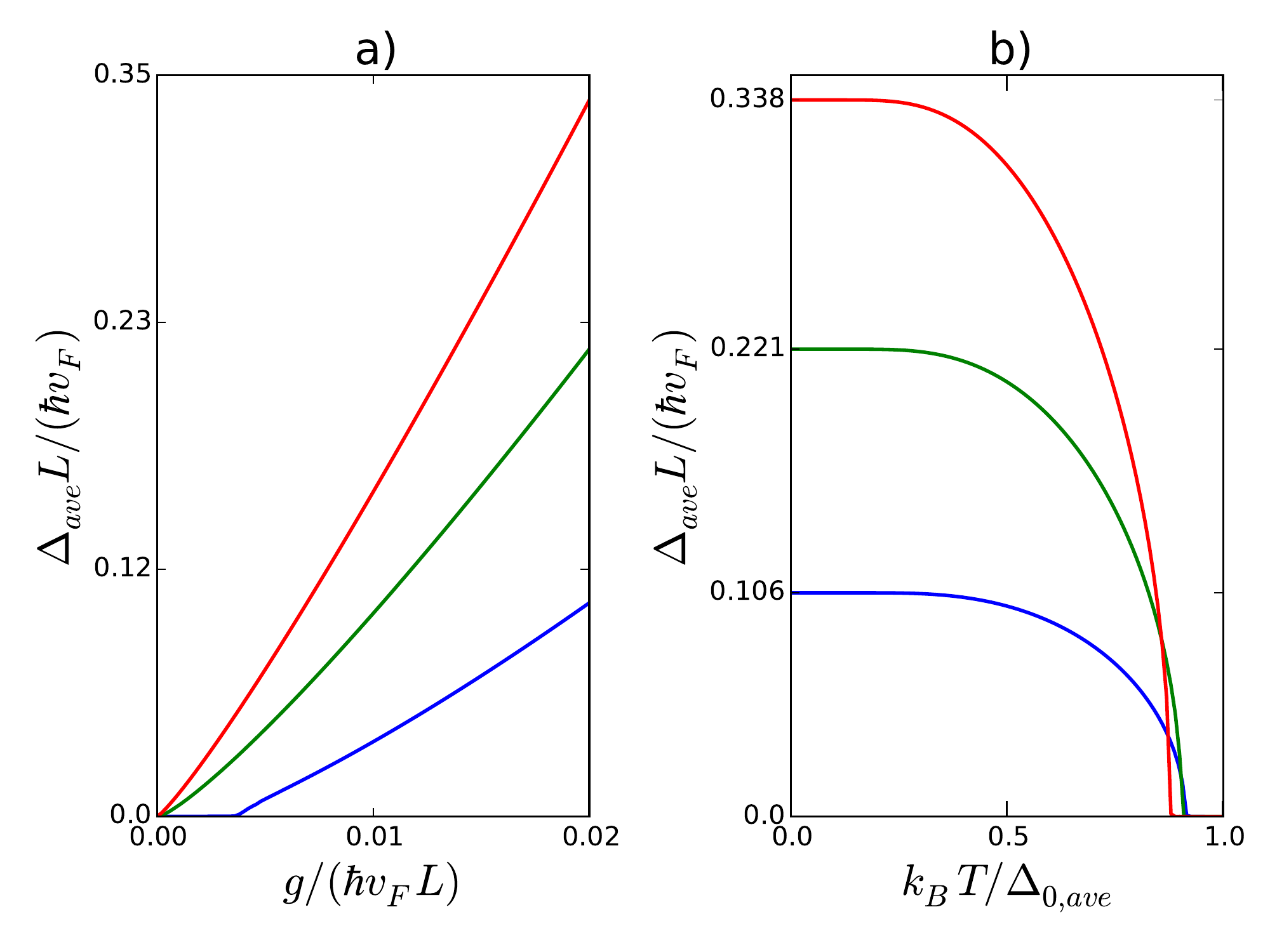}
\caption{\small a) Dependence of $\Delta_{\rm ave}$ on superconducting coupling with $\beta=20$ (blue), $\beta=30$ (green) and $\beta=40$ (red). b) $\Delta_{\rm ave}$ as a function of temperature for $g / (\hbar v_F L) =0.02$.}
\label{fig:delta}
\end{figure}

In Fig.~\ref{fig:delta}b we also plot the average of $\Delta(x)$ as a function of the coupling constant. For small couplings, where $\Delta$ is small and can ``see'' the small linear slope of the spectrum, there is a quantum critical point at $g \approx \pi^2 \hbar c L / \beta$ below which there is no superconductivity. In the figure, this point is only visible for $\beta=20$ because for larger $\beta$ the linear regime of the spectrum becomes exponentially smaller. For larger $g$, the system enters the flat band regime. A simple constant $\Delta$ estimate yields for the order of magnitude estimate $\Delta \approx \beta g / (2 L^2 )$. This expression also shows the strong linear relationship between $\Delta$ and $g$ that is apparent in the numerical calculation. From the numerics, we can also find that the critical temperature $T_c$ is approximately given by the average value of $\Delta$, i.e. $k_B T_c \approx (1/L) \int_0^L dx \Delta(x)$. This behaviour is shown in Fig.~\ref{fig:delta}b. Due to the linear dependence between $T_c$ and $g$ we can therefore expect a high critical temperature in this parameter regime.

We also calculate the spectrum and local density of states in the superconducting state. At zero momentum, the spectrum has a gap of magnitude $\vert E_{\mathbf{p}=0} \vert \approx 2\Delta_{max}$ which is the expected result for any superconductor. However, for small momenta, the slope of the spectrum is negative for excitations with positive momentum and energy, leading to a local minimum at $k_y = k_{\rm min} \sim \beta / (2 L)$. The momentum dependence results from the localized $\Delta(x)$. In the density of states, the two local minima of the spectrum lead to a peculiar two-peaked shape shown in Fig.~\ref{fig:dos}a. This feature could act as a possible experimental signature for superconductivity described by this model.

For a flat band, the group velocity $c$ of the electrons becomes very small for both normal and superconducting state (see Fig.~\ref{fig:dos}d). It would hence be natural to think of the paired electrons to be localized, unable to carry supercurrent. However, there are also other contributions to supercurrent besides those proportional to the group velocity \cite{peotta2015}. We calculate the supercurrent by adding a phase gradient to the order parameter as $\Delta(x) \rightarrow \Delta(x) e^{i \mathbf{k}_s x}$ in which case the supercurrent is given by
\begin{equation}
\mathbf{J}(x) =\frac{e v_F}{L} \sum_{\mathbf{p},n} \left(f_{\mathbf{p},n} u^\dagger_{\mathbf{p},n} \mathbf{\sigma} u_{\mathbf{p},n} + (f_{\mathbf{p},n}-1) v^\dagger_{\mathbf{p},n} \mathbf{\sigma} v_{\mathbf{p},n}\right),
\end{equation}
where $f_{{\mathbf p},n}$ is the number of quasiparticles occupying the $n$th band at momentum ${\mathbf p}$. At $T=0$, it is $f[E_n({\mathbf p})]=1$ for $E_n({\mathbf p})<0$ and 0 otherwise.

\begin{figure}[!ht]
\centering
\includegraphics[width=0.98\columnwidth]{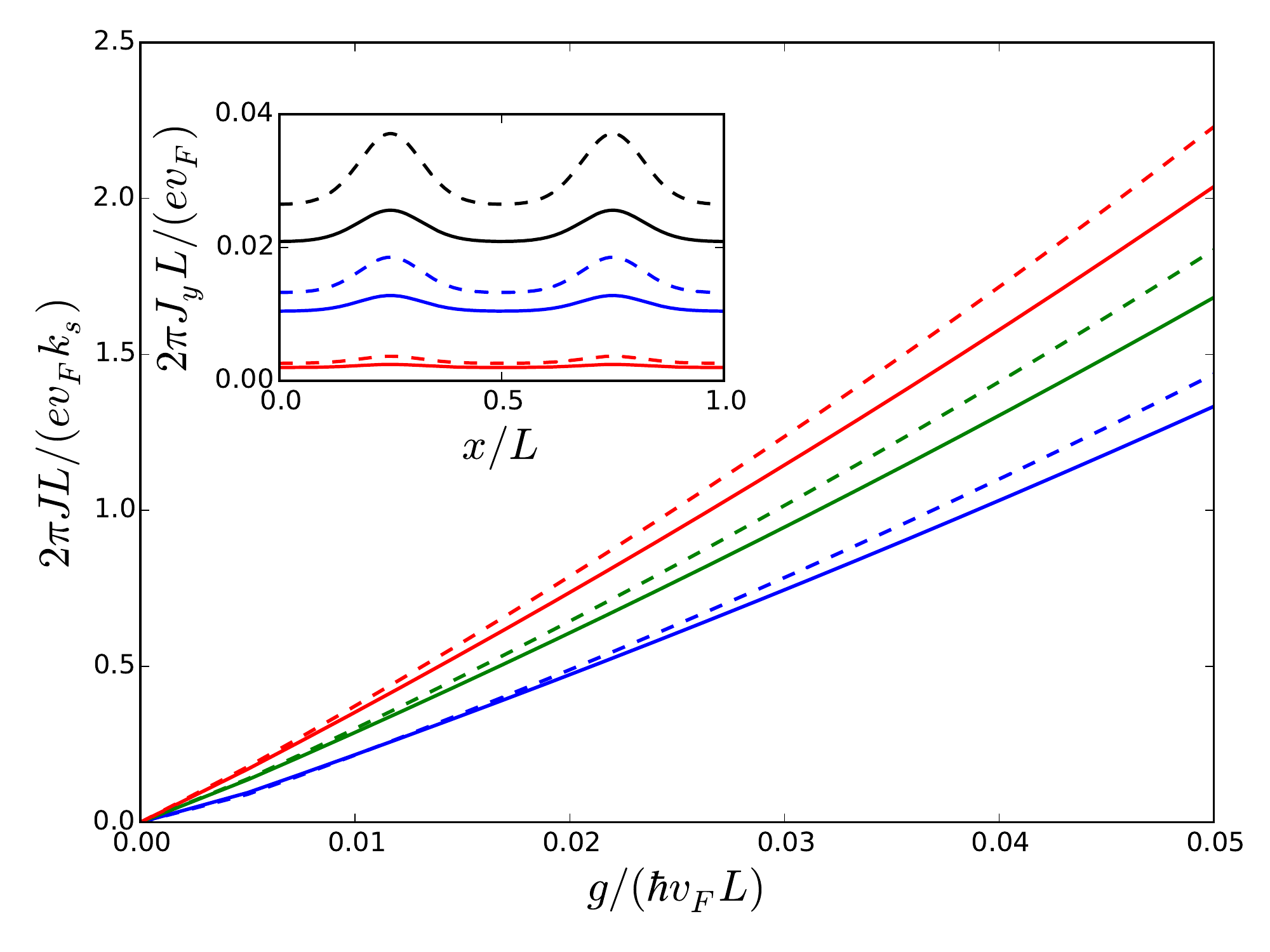}
\caption{Supercurrent in $x$ and $y$ directions (solid and dashed lines, respectively) as a function of the superconducting coupling for $\beta=25$ (blue), $\beta=30$ (green) and $\beta=35$ (red). Inset: profile of the supercurrent in $y$ direction for $\beta=25$ (solid line) and $\beta=30$ (dashed line) for $k_s L =0.01, 0.05, 0.1$.}
\label{fig:sc}
\end{figure}

For $\vert \mathbf{k}_s L \vert \ll 1$, we find that the supercurrent is approximately of the form $J_i(x) \approx a_i \Delta_{max} k_{s,i}$, where $i \in \{x,y\}$ and $a_x = 0.17$, $a_y = 0.19$ are fitted constants that describe the weak anisotropy of the current. This result is shown in Fig.~\ref{fig:sc}. While current in the $x$ direction must be independent of $y$ due to translational invariance, it can be inhomogenous in the $y$ direction. In the inset of Fig.~\ref{fig:sc} we show the profile of current flowing in $y$ direction. Interestingly, it is only weakly inhomogenous even though the superconducting order parameter varies strongly in space. The reason for this is that the current is proportional to the overlap of the wave functions of the two sublattices and the overlap is almost position independent. This behaviour is analogous to what happens in a model for superconducting rhombohedral graphite \cite{kopnin2011surface}.

Let us discuss the possible physical realizations of periodically strained Dirac fermions. So far, graphene has been shown to be superconducting when doped with calcium and possibly also with other elements \cite{chapman2015superconductivity,profeta2012phonon,fedorov2014observation}. Doping of the pure Dirac material is required to increase the density of states so that the quantum critical point disappears. If this is the main effect, then the scheme of a periodic strain discussed here should also make even undoped graphene superconducting because the flattening of the spectrum greatly lowers the superconducting coupling required to overcome the quantum critical point. Contrary to the pristine graphene (or other Dirac material), doping the strained graphene to move the chemical potential out of the flat band regime would act to reduce the critical temperature. 

Strain superlattice could explain the observations of the superconducting-type behavior at interfaces between graphite regions with different lattice orientations \cite{scheike2012can,ballestar2013josephson,esquinazi2008indications}. It is possible that such interfaces stabilize an array of screw dislocations, which would lead to the presence of periodically strained graphene at the interfaces. 

We can estimate the value of the critical temperature in periodically strained graphene using the coupling constant obtained from experiments \cite{fedorov2014observation,chapman2015superconductivity}. As shown in \cite{supplement}, we would get $T_c$ as large as 420 K for $\beta=30$ and $L=10$ nm. In this estimate, we neglect the effects from the strain superlattice or the doping in the experiments on $g$, and therefore it should be taken only as indicative. Such effects are left for further work.

Another class of materials where flat band superconductivity due to straining has already been suggested is layered structures made out of IV-VI semiconductors such as PbTe/SnTe, PbSe, PbS, PbTe/PbSe, PbS/YbS and PbTe/YbS \cite{fogel2002interfacial,fogel2001novel}. Our results here could be used to verify whether models previously suggested \cite{tang2014strain} indeed are valid in these materials.

Besides superconductivity, flat bands can promote also other types of states, such as magnetism, depending on the dominant interaction channel. Moreover, in such two-dimensional systems the long-range correlations are most likely suppressed by some mechanism, which would limit the observation of superconductivity in large samples. Besides the phase fluctuations leading to Kosterlitz-Thouless physics, another mechanism for suppressing correlations would be those affecting the strain lattice. Therefore, on length scales long compared to such elastic correlation length, the system would most likely be described by a set of Josephson coupled superconducting islands. The exact elastic correlation lengths are materials dependent and therefore out of the scope of the present work. We also note that in the case of strained graphene, other pairing symmetries have been considered \cite{roy2014strain}.

We thank Grigori Volovik and Pablo Esquinazi for fruitful discussions. This work was supported by the Academy of Finland through its Center of Excellence program project number 284594, and by the European Research Council (Grant No. 240362-Heattronics).

\end{document}